\documentclass[journal,letterpaper,12pt,oneside,onecolumn,draftclsnofoot]{IEEEtran}
\makeatletter
\def\ps@headings{%
\def\@oddhead{\mbox{}\scriptsize\rightmark \hfil \thepage}%
\def\@evenhead{\scriptsize\thepage \hfil \leftmark\mbox{}}%
\def\@oddfoot{}%
\def\@evenfoot{}}
\makeatother

\usepackage{fancyhdr}
\usepackage{graphicx}
\usepackage{subfigure}
\usepackage{amsmath}
\usepackage{epstopdf}
\usepackage{booktabs}
\usepackage[english]{babel}
\usepackage{amssymb}

\usepackage[ruled,vlined]{algorithm2e}
\usepackage{tipa}
\usepackage{lpic}
\usepackage{stfloats}
\usepackage{algorithmic}
\usepackage{url}
\usepackage{caption2}
\usepackage{verbatim}
\usepackage{empheq}
\usepackage{amsthm}
\usepackage{cases}
\usepackage{multirow}
\usepackage{longtable}
\usepackage{tabu}

\ifCLASSINFOpdf

\else

\fi

\begin{document}

\title{Vehicular Communications: A Network Layer Perspective}

\author{\IEEEauthorblockN{Haixia Peng,~\IEEEmembership{Member,~IEEE}, Le Liang,~\IEEEmembership{Student Member,~IEEE}, Xuemin (Sherman) Shen,~\IEEEmembership{Fellow,~IEEE}, Geoffrey Ye Li,~\IEEEmembership{Fellow,~IEEE}}

\thanks{
H. Peng and X. (S.) Shen are with the Department of Electrical and Computer Engineering, University of Waterloo, Waterloo, ON, Canada, N2L 3G1 (e-mail: h27peng@uwaterloo.ca, sshen@uwaterloo.ca).
L. Liang and G. Y. Li are with the School of Electrical and Computer Engineering, Georgia Institute of Technology, Atlanta, GA (e-mail: lliang@gatech.edu; liye@ece.gatech.edu).}}

\maketitle
\cfoot{\thepage}
\renewcommand{\headrulewidth}{0pt}
\renewcommand{\footrulewidth}{0pt}
\pagestyle{fancy}
\cfoot{\thepage}

\begin{abstract}

Vehicular communications, referring to information exchange among vehicles, pedestrians, and infrastructures, have become very popular and  been widely studied recently due to its great potential to support intelligent transportation and various safety applications. Via vehicular communications, manually driving vehicles and autonomous vehicles can collect useful information to improve traffic safety and support infotainment services. In this paper, we provide a comprehensive overview of recent research on enabling efficient vehicular communications from the network layer perspective. First, we introduce general applications and unique characteristics of vehicular networks and the corresponding classifications. Based on different driving patterns of vehicles, we divide vehicular networks into two categories, i.e., manually driving vehicular networks and automated driving vehicular networks, and then discuss the available communication techniques, network structures, routing protocols, and handoff strategies applied in these vehicular networks. Finally, we identify the challenges confronted by the current vehicular communications and present the corresponding research opportunities.

\end{abstract}

\begin{IEEEkeywords}
Vehicular communications; manually driving vehicles; autonomous vehicles; routing protocols; handoff strategies; communication technologies.
\end{IEEEkeywords}

\section{Introduction}
\label{sec:Intro}

In the last decade, vehicular communications have attracted tremendous interest from both academia and industry. We have provided a comprehensive survey on vehicular communications from the perspective of the physical layer in \cite{liang2017vehicular}. Via enabling vehicles to exchange information with other vehicles (i.e., vehicle-to-vehicle, V2V), pedestrians (i.e., vehicle-to-pedestrian, V2P), and infrastructures (i.e., vehicle-to-infrastructure,
V2I), vehicular communications are expected to support a variety of applications, including intelligent transportation and safety applications \cite{sichitiu2008inter, ma2012design}. However, due to the high mobility and complicated communications environments, it is very challenging to provide efficient vehicular communications to satisfy the different requirements, especially, higher reliability and lower latency for sharing safety-related information.

Manually driving vehicular networks (MDVNETs), referring to communications among vehicles with only manually driving patterns, are one of the main applications of vehicular communications and have been widely considered in the existing works \cite{azees2016comprehensive}. MDVNETs can help manually driving vehicles to improve the traffic safety and provide infotainment services to drivers and passengers \cite{ma2012design}.

Owing to the advances in sensor technologies, wireless communications, computational power, and intelligent control, a new driving pattern, named automated driving, is gradually applied in vehicles, and then autonomous vehicles (AVs) are generated and begin to be test-driven on the roads. According to a report from the National Highway Safety Administration (NHTSA) \cite{national2013preliminary}, passenger vehicles can be classified into five distinct levels of autonomy, as shown in Table \ref{table:levels}. Manually driving vehicles mentioned in this paper refer to the vehicles with level 0 autonomy while AVs refer to vehicles with level 4 autonomy. It is expected that no human actions or interventions are required for an AV that can automatically navigate a variety of environments, other than setting the destination and starting the system. That is, people can be relieved from the stress of driving \cite{koopman2017autonomous, blyth2016expanding}. Moreover, AVs have been indicated as a good solution to many traffic related issues, such as traffic accidents, traffic congestion, exhaust pollution, and fossil fuels overusing. Current vision of AVs promises a future where increased safety, velocity, convenience, and comfort are offered while energy consumption is reduced \cite{blyth2016expanding}.

\begin{table}[htbp]
\begin{center}
\setlength{\belowcaptionskip}{5pt}
\caption{\label{table:levels} Five distinct autonomy levels of passenger vehicles}
\centering
\renewcommand\arraystretch{1.8}
\begin{tabu} to 1 \textwidth{X[0.2,l]||X[0.8,l]}
   \toprule
    Autonomy levels & Description \\
    \midrule
    0 - No automation & The primary vehicle controls, such as throttling, braking, and steering, are completely handled by the driver at all times. \\
    1 - Function specific automation & The primary vehicle controls are mainly handed by the driver while one or more specific control functions, such as electronic stability control, are involved to help control the vehicle.\\
    2 - Combined function automation & Two or more primary control functions are involved to help control the vehicle to relieve the driver of controlling of those functions. Adaptive cruise control in combination with lane centering is one of examples of combined functions. \\
    3 - Limited self-driving automation & All safety-critical functions are automated under certain traffic environments while the vehicle should monitor the environment to transit back to driver control. \\
    4 - Full self-driving automation & All safety-critical driving functions and monitor roadway conditions for an entire trip are performed by the vehicle and drivers only need to provide destination or navigation input. \\
    \bottomrule
\end{tabu}
\end{center}
\end{table}

Despite the above attractive advantages of AVs, how to ensure the autonomous driving system to be safe enough to leave humans' actions and interventions completely presents significant challenges. Being ``safe" means at least the AVs can correctly execute vehicle-level behaviors, such as obeying traffic regulations and dealing with road and roadside hazards \cite{hobert2015enhancements, koopman2017autonomous}. Plenty of works have indicated that inter-vehicle information plays an important role for AVs' ``safe", in which, helpful information, either via onboard sensors or vehicular communications, is needed for the cooperative driving among AVs and collision avoidance \cite{moradi2016hybrid}, and therefore enabling advanced features, such as ``platooning", and alerting AVs of real-time mapping information and surrounding environments, such as other AVs and potential hazards \cite{Nguyen2017region}. Thus, automated driving vehicular networks (ADVNETs), used for wireless communications among AVs, have been regarded as another important application of vehicular networks.

Existing studies and testings have indicated that there are several potential wireless communication technologies that can be applied to support vehicular communications either in MDVNETs or ADVNETs: dedicated short-range communications (DSRC) \cite{biswas2006vehicle}, cellular network technologies \cite{liang2017resource, Abboud2016Interworking, chen2016lte}, Wi-Fi \cite{lu2016wi}, White-Fi \cite{zhou2016toward}, infrared (IR) \cite{fernandes2012platooning}, and visible light communications (VLC) \cite{pathak2015visible}. Through using one of these technologies or their combination, a variety of data generated by vehicles can be shared successfully to support different applications. However, in addition to the fundamental physical layer issues discussed in our previous work \cite{liang2017vehicular}, there are many network layer issues that should be taken into account to enable efficient vehicular communications. Different types of communications techniques may fit in different vehicles in different environments. How to specify the ways that different communication entities exchange information with each other is a key to achieve efficient message dissemination, which refers to designing efficient routing protocols for different messages \cite{li2007routing}. How to select network for executing information sharing when two or more technologies are available, i.e., developing corresponding handoff strategy, will determine the general network performance, such as delay \cite{zhu2011mobility}.

Over the past decade, there have been some excellent survey papers on vehicular communications from a network layer perspective, such as \cite{li2007routing, zhu2011mobility, cooper2017comparative, Abboud2016Interworking}, focusing on the network layer issues in MDVNETs. In this paper, we present a comprehensive overview of recent research on enabling efficient vehicular communications both in MDVNETs and ADVNETs. We will start with general applications and unique characteristics of vehicular networks and the corresponding classifications. Then, we discuss recent research from a network layer perspective for both MDVNETs and ADVNETs. Specifically, for MDVNETs, we analyze the advantages and disadvantages of different communication technologies and introduce the routing protocols and handoff strategies. For ADVNETs, we discuss two short-range communication technologies, i.e., IR and VLC, and then summarize the communication structures designed for ADVNETs with different traffic management strategies. This survey also helps identify the main challenges and potential solutions for the coming application scenario, in which communications may be needed between manually driving vehicles and AVs or among vehicles with level 3 autonomy.

The rest of this paper is organized as follows. In Section \ref{sec:General_Veh}, the general applications and unique characteristics of vehicular networks are introduced. In Section \ref{sec:Indivi}, we first analyze four different communication technologies and then address the routing and handoff issues for MDVNETs. Two short-range communication technologies and communication structures that can be applied in ADVNETs are analyzed in Section \ref{sec:Platoon}. Then in Section \ref{sec:CAO}, we identify the main challenges and present the potential solutions. Finally, we conclude this paper in Section \ref{sec:Conclu}.

\section{General Vehicular Networks}
\label{sec:General_Veh}

In this section, we will discuss general vehicular networks, which are usually for data exchange among vehicles without considering the driving pattern of vehicles. We will first introduce the applications of vehicular networks and then describe the unique characteristics of vehicular networks. Note that in-vehicle communications that refer to the wired or wireless communications between an on-board unit (OBU) and one or multiple application units (AUs) in a vehicle \cite{zeng2016vehicle}, are not considered here. Moreover, the mentioned AVs should be distinguished from autonomous robots, unmanned aerial vehicles, and unmanned underwater vehicles even if many communication technologies introduced here can be directly applied there.

\subsection{Vehicular Networks Applications}
\label{subsec:VNA}

For moving vehicles, communications networks are usually designed for sharing information and supporting a large number of cooperative applications.

\begin{figure}[htbp]
\centering
\includegraphics[height=2.2 in]{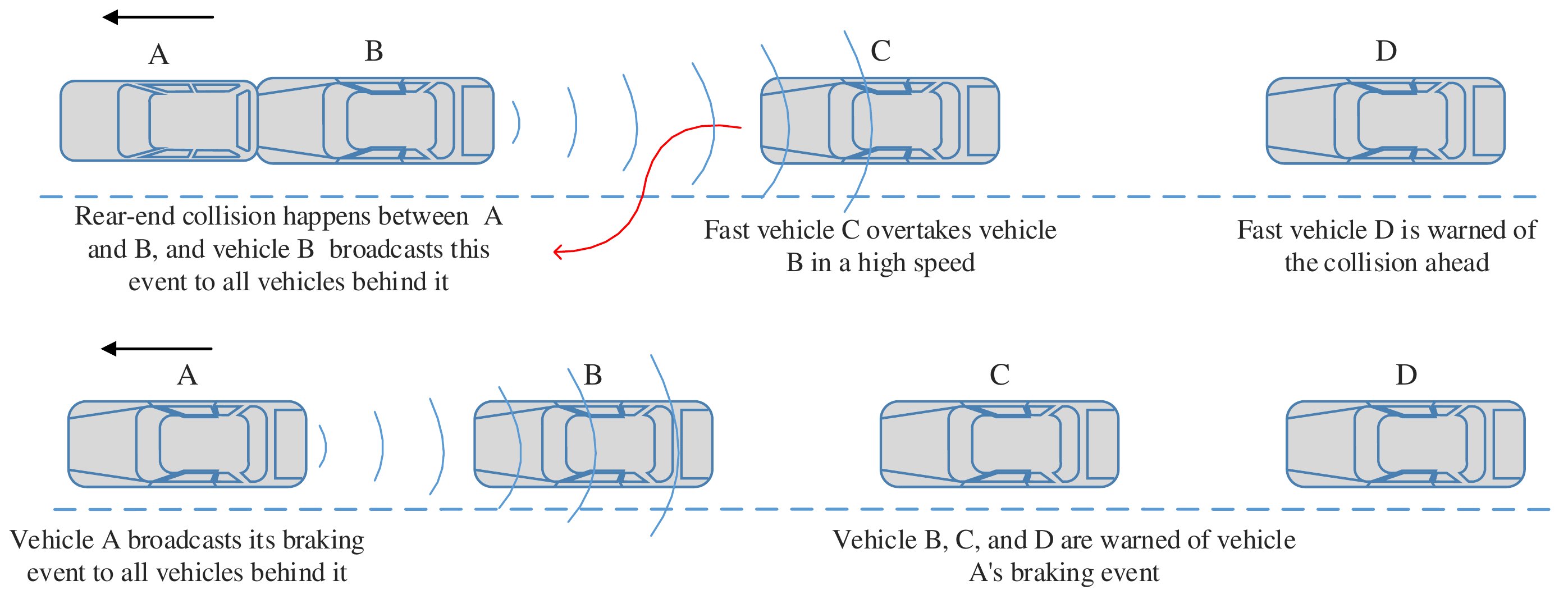}
\caption{Examples for sharing event-driven safety information}
\label{Fig.Rear-end_collision}
\end{figure}

\textbf{\emph{Safety applications:}} Through sharing safety-related information \cite{sichitiu2008inter}, safety services can be provided, the traffic accidents can be significantly reduced, and the commuters' life, health, and property can be effectively protected. Once obtaining safety-related information from other vehicles, drivers will take actions in advance to enhance driving safety or be informed about expected dangerous situations in advance to avoid traffic accidents \cite{azees2016comprehensive}. One type of safety-related information is vehicle's traveling state information, such as, current position, real-time speed, and direction. This type information is not only important to assist the drivers or the automated driving systems in passing and changing lane and avoiding collision, but also a necessary condition for platoon-based driving with autonomous vehicles to maintain the string stability of platoons \cite{xu2014communication}. Another type of safety-information is event-driven safety information, e.g., emergency vehicle warning, traffic condition warning, emergency electronic brake lights, cooperation collision warning, and rear-end collision warning. Event-driven safety information, generated by certain vehicles involved in or discovering a dangerous situation, such as an emergency brake or sudden lane change, should be shared to help other vehicles obtain real-time situational awareness and detect possible dangers. As shown in Figure \ref{Fig.Rear-end_collision}, sharing cooperation collision and rear-end collision warning information among vehicles can help avoid accidents in several scenarios \cite{jiang2006design}.


\textbf{\emph{Non-safety applications:}} By sharing information among moving vehicles, value-added services, such as traffic management and infotainment support can be provided, to enhance the comfort of the commuters. Similar to some safety applications, most of the traffic management applications are designed for reducing traffic jams to improve traffic flow and save travel time for commuters. For example, via sharing information about traffic monitoring and road conditions among moving vehicles, traffic management applications can be applied to help drivers reroute to their destinations and to improve the efficiency of traffic light schedules, and consequently reduce traffic jams \cite{hernandez2014reliable}. Another common example of traffic management applications is electronic toll collection. Via helping the toll gates obtain the information about vehicle identification, electronic toll collection can eliminate the delay on collecting tolls to save travel time for commuters \cite{ETC}. Different from traffic management, infotainment support applications are mainly focusing on providing traveler location-based services and entertainment. For instance, infotainment support applications can provide location information, such as about fuel station, parking, restaurant, and hotel, to moving vehicles when the related services are required by the drivers or passengers \cite{zeadally2012vehicular}. Furthermore, infotainment support applications can also provide access to the Internet for the moving vehicles to download multimedia entertainment information \cite{Lee2014Khop}.

Despite the benefits of the above cooperative applications, the implementation of safety and non-safety applications has their own challenges. In a vehicular network with certain wireless communication technology, all available spectrum resources can be used for either safety or non-safety applications. For safety applications, related information is with high priority in terms of transmission delay and reliability so that the driver can receive them and take the corresponding actions in time \cite{zeadally2012vehicular}. Moreover, due to the high mobility, variable network density, and unstable topology in vehicular networks, meeting these requirements is sometimes very challenging \cite{jiang2006design, biswas2006vehicle, ma2012design, ucar2016multihop, omar2016wireless}. It is for these reasons that safety applications are given higher priority over non-safety applications. Different from safety applications, most of the non-safety applications do not have stringent real-time requirements. However, how to reduce the delay and packet loss for non-safety information without impacting safety applications is important to improve the service quality of the non-safety applications, especially for some infotainment applications \cite{Lee2014Khop, zaimi2016improved}.

\subsection{Vehicular Networks Characteristics}
\label{subsec:VCA}

In vehicular networks, wireless communications are basically executed by applying the principles of mobile ad hoc networks (MANETs), namely, wireless communications are spontaneously created for data exchange. In addition to some characteristics similar to MANETs, such as self organization and management, short to medium transmission range, omnidirectional broadcast, and low bandwidth, a vehicular network has its own unique characteristics due to its moving nodes. According to whether they are beneficial to information exchange for vehicles or not, these characteristics are classified into two categories, detrimental and beneficial characteristics.

\textbf{\emph{Detrimental characteristics:}} These characteristics pose obstacles or challenges to communications over vehicular networks, including high mobility, stringent delay constraints, and complicated communication environments.

\emph{1) High mobility:} Due to the high moving speed of vehicles, vehicular networks even have higher mobility than MANETs \cite{cheng2007mobile}. High mobility often results in frequently disconnected wireless communication links and then reduces the effective communication time among vehicles. Furthermore, it also causes the network topology to change dynamically and further adds challenge to information exchange among vehicles.

\emph{2) Stringent delay constraints \cite{li2007routing}:} In some vehicular network applications, such as safety applications and some infotainment applications, information exchange is required to be successfully finished within a particular time limit to avoid traffic accidents, protect the commuters' life, and ensure the quality of infotainment services. Note that delay mentioned here is the maximum delay from the source to the destination, not average delay in vehicular networks.

\emph{3) Complicated communications environments:} Vehicular networks are usually applied in three communications environments. The first one is one-dimensional communications environments, such as highway traffic scenario. Even though vehicles in highways always move faster than in other environments, this environment is relatively simpler due to the straightforward moving direction and relatively fixed speeds. The second environment is two-dimensional communications environments. A typical example is urban traffic scenarios, which are more complex compared with the highway scenario \cite{togou2016scrp}. The streets in most urban areas are often divided into many segments at the intersections. Therefore, for two moving vehicles in different road segments, a direct communication link may not exist due to the obstacles around the intersections, such as buildings and trees. Moreover, there are always higher traffic density of vehicles in urban areas, which implies more communication links within communication range and significantly impacts the spectrum resource occupation probability. The last one is three-dimensional communications environments, such as viaducts \cite{zhu2016geographic}. For vehicles in a viaduct, communication links in different physical space layers make this environment the most complex.

\textbf{\emph{Beneficial characteristics:}} These characteristics are beneficial to the wireless communications in vehicular networks, such as weak energy constraints and driving route prediction.

\emph{1) Weak energy constraints:} Since vehicles always have sufficient energy, vehicular networks do not suffer from power constraints as in regular mobile communications networks. Moreover, each vehicle can afford significant sensing and computing capabilities, including data storage and processing since it can power itself while providing continuous communication with other vehicles.

\emph{2) Driving route prediction:} Vehicles in the networks are limited to moving on the road in usual circumstances, which makes it possible to predict the driving route for itself or even for other vehicles when the road map and vehicle speed information are available. Driving route prediction plays an important role in routing protocol design for vehicular networks, especially when addressing the challenges presented by the high mobility.

Characteristics of vehicular networks are similar no matter in MDVNETs or ADVNETs. However, the challenges presented by the above characteristics and how to address these challenges to meet the requirements of the applications are different for different types of vehicular networks. We will address these issues from the network layer perspective in the subsequent three sections.

\section{Manually Driving Vehicular Networks}
\label{sec:Indivi}

Almost all current vehicles are manually driving and always move individually on the roads. For example, some drivers may accelerate suddenly to pass other vehicles or other drivers may get used to a low speed. Thus, the impacts of the high and heterogeneous mobility on the manually driving vehicular networks are significant. How to address the challenges caused by high and heterogeneous mobility and enable communications in different environments has attracted more and more attention. In this section, we will discuss the existing works for MDVNETs. Specifically, we first introduce the available communication technologies for MDVNETs and then present routing protocols and handoff strategies.

\subsection{Technologies for MDVNETs}
\label{subsec:ACT}

In addition to navigation system, most modern vehicles are fitted with DSRC, cellular, Wi-Fi, White-Fi etc., to enable the vehicular networks to improve the driving experience and safety \cite{alsath2014shared}. In the following, we review four widely applied communication technologies in MDVNETs. Note that, ultra-wide band (UWB) and bluetooth technology, which can be used to support the in-vehicle communications or/and traffic conditions monitoring \cite{bas2013ultra, laharotte2015spatiotemporal}, are not considered here.

\textbf{\emph{DSRC:}} It is a dedicated wireless communication technology used for information exchange among moving vehicles over short to medium range \cite{biswas2006vehicle}. As the only communication technology specifically designed for vehicular users, DSRC can provide
\begin{enumerate}
\item Designated licensed bandwidth: In October 1999, 75 MHz radio spectrum in the 5.9 GHz band was allocated to support the DSRC-based communications in intelligent transportation system (ITS) applications by the Federal Communications Commission (FCC) of the United States;
\item High reliability: DSRC-based wireless links can work in high mobility and harsh weather conditions, such as rain, fog, and snow;
\item Priority for safety applications: The total 75 MHz bandwidth is divided into one control channel (CCH) and six service channels (SCHs), each with 10 MHz bandwidth and 5 MHz guard band. Among these seven channels, safety applications are given priority over non-safety applications \cite{kenney2011dedicated};
\item Security and privacy: In DSRC, message authentication and privacy are provided by the IEEE 1609.2 standard \cite{its2013ieee}.
\end{enumerate}

Thanks to the above benefits, DSRC is regarded as one of the most promising technologies applied to support ITS applications, especially the safety-related ones \cite{cheng2007mobile}. However, DSRC also exposes some drawbacks. First, due to the limited spectrum resource, broadcast storm may occur when disseminating safety information over a large area, especially in the situation with high vehicle traffic density. With the increase in the number of vehicles attempting to transmit in the same channel simultaneously, the packet delay and transmission collisions probability will be increased and the performance of DSRC will degrade \cite{ucar2016multihop, omar2013vemac}. To address this problem, improved broadcast mechanisms, such as probabilistic flooding and clustering \cite{ucar2016multihop, wu2015efficient}, have been proposed and will be introduced in detail in the next subsection. Another obvious drawback in DSRC communications is poor and short-lived connectivity, including V2V connectivity and communications between a vehicle and a road-side unit (RSU), i.e., V2R connectivity. Short-lived V2V connectivity always occurs in the environment with low vehicle density, where the number of vehicles is too sparse to disseminate the information to all the destination vehicles. Furthermore, due to the short radio transmission distance, i.e., around 300 m, and DSRC can only provide short-lived V2R connectivity \cite{araniti2013lte} if there is no pervasive roadside communication infrastructure. In order to improve the performance of vehicular networks, some medium and long-rang communication technologies \cite{ucar2016multihop} can be commonly used.

\textbf{\emph{Cellular:}} Nowadays, cellular networks are distributed over land areas, where each cell is served by a base station (BS), e.g., the evolved node B (eNB) in the long-term evolution (LTE) system. The key enabler of cellular-based vehicular networks is the LTE standard developed by the 3rd Generation Partnership Project (3GPP), which provides efficient information dissemination to many user equipment \cite{ucar2016multihop}. Lots of academic research and field tests have indicated that cellular technologies, such as the worldwide interoperability for microwave access (WiMAX) and (LTE) wireless technologies, possess certain advantages in vehicular networks \cite{araniti2013lte, rengaraju2014qos}. Specifically, benefited from the large coverage area of the eNB and high penetration rate, cellular technologies can provide relatively long lived connectivity to support V2I communications \cite{araniti2013lte}. Compared with other wireless communication technologies, cellular technologies can potentially support several vehicle users within a small region simultaneously due to its relatively high capacity. Furthermore, the channel and transport modes in cellular technologies, i.e., the dedicated/common modes and the unicast/broadcast/multicast downlink transport modes, can help reduce the transmission delay and improve the capacity for communication environments with high vehicle density \cite{ucar2016multihop}. Device-to-device (D2D) communications can provide short range direct links between two vehicle users to reuse the spectrum, and therefore mitigate the problems caused by the limited radio spectrum resources \cite{su2016d2d, liang2017resource, cheng2017performance}.

In the last few years, cellular-based vehicular networks have been widely investigated \cite{araniti2013lte, liang2017resource}. Due to the above mentioned advantages, cellular technologies are regarded as a promising alternative to DSRC for vehicular networks \cite{araniti2013lte}. However, due to the current cellular data pricing model, the corresponding cost for data transmission in cellular-based vehicular networks is much higher than other wireless communication technologies \cite{lu2016wi}. On the other hand, in the dense traffic areas, the heavy data traffic-load generated by some vehicles may significantly challenge the cellular capacity and potentially affect the delivery of the traditional cellular applications \cite{araniti2013lte}. To address this challenge, millimeter-wave (mmWave) communications with advantages of multi-gigabit transmit ability and beamforming technique have been considered for the 5th generation (5G) cellular networks. For example, millimeter-wave communications are applied for sharing vehicles' massive sensing data in \cite{choi2016millimeter}, where the beam alignment overhead has been reduced by configuring the mmWave communication links based on sensed or DSRC-based information.

\textbf{\emph{Wi-Fi:}} It is a technology for wireless local area networks (WLANs) based on the IEEE 802.11 standards. It has been shown in \cite{lu2016wi,lu2013vehicles,taleb2015vecos,xu2017delay} that Wi-Fi technology is an attractive and complementary Internet access method for moving vehicles. Equipped with a Wi-Fi radio or Wi-Fi-enabled mobile devices, such as a mobile phone, vehicles can access the Internet when they drive through the coverage of Wi-Fi access points. The obvious advantages of Wi-Fi technology include low per-bit cost, extremely wide spread global deployments, and higher peak throughput, which are beneficial to some vehicular applications with a high data transmission rate, such as infotainment applications. However, due to the limited coverage of each Wi-Fi access point (AP) and the high mobility of vehicles, Wi-Fi technology suffers from intermittent connectivity in vehicular networks \cite{lu2013vehicles}. Thus, handoff schemes become particularly important to Wi-Fi technology in such a scenario \cite{lv2015swimming}. Furthermore, instead of establishing the Wi-Fi-based vehicular networks for the inter-vehicle communications, Wi-Fi technology is considered as an complementary access method to offload delay tolerant data traffic \cite{lu2016wi,cheng2016opportunistic,taleb2015vecos}.

\textbf{\emph{White-Fi:}} It is a term coined by the FCC of the United States to describe communications that allows unlicensed users to access the TV white space spectrum in the VHF/UHF bands between 54 and 790 MHz. Note that even though White-Fi is also referred to super Wi-Fi, it is not endorsed by the Wi-Fi Alliance or based on Wi-Fi technology. The progress of White-Fi technology has yielded many new insights into vehicular networks, which has motivated researchers to explore unlicensed spectrum to solve the spectrum-scarcity issue for vehicular networks. It has been shown in \cite{lim2014interplay} and \cite{zhou2016toward} that the White-Fi-enabled vehicular networks can improve the dissemination capacity of vehicles by offloading a portion of data traffic from the DSRC band or cellular band to the TV band. Furthermore, different from the 2.4 GHz radio frequency used by Wi-Fi, TV white space spectrum is at lower frequency range and  allows the signal to penetrate walls better and travel further than the higher frequency range. Thus, White-Fi technology can provide relatively long rang communications to improve the transmission efficiency. For example, applying White-Fi for long-distance dissemination to avoid multi-hop transmission can reduce the delay of some safety related information \cite{lim2014interplay}. However, White-Fi-enabled vehicular communications are potential interference to incumbent TV band users, which may bring challenges to protect the incumbent service. Moreover, due to the unlicensed characteristic of TV band, vehicular networks and other existing wireless networks are all allowed to use. Vehicle users may experience interference caused by other secondary networks, and therefore impact the service quality of some vehicular applications \cite{han2017vehicular}.

\begin{figure}[htbp]
\centering
\includegraphics[height=3.4 in]{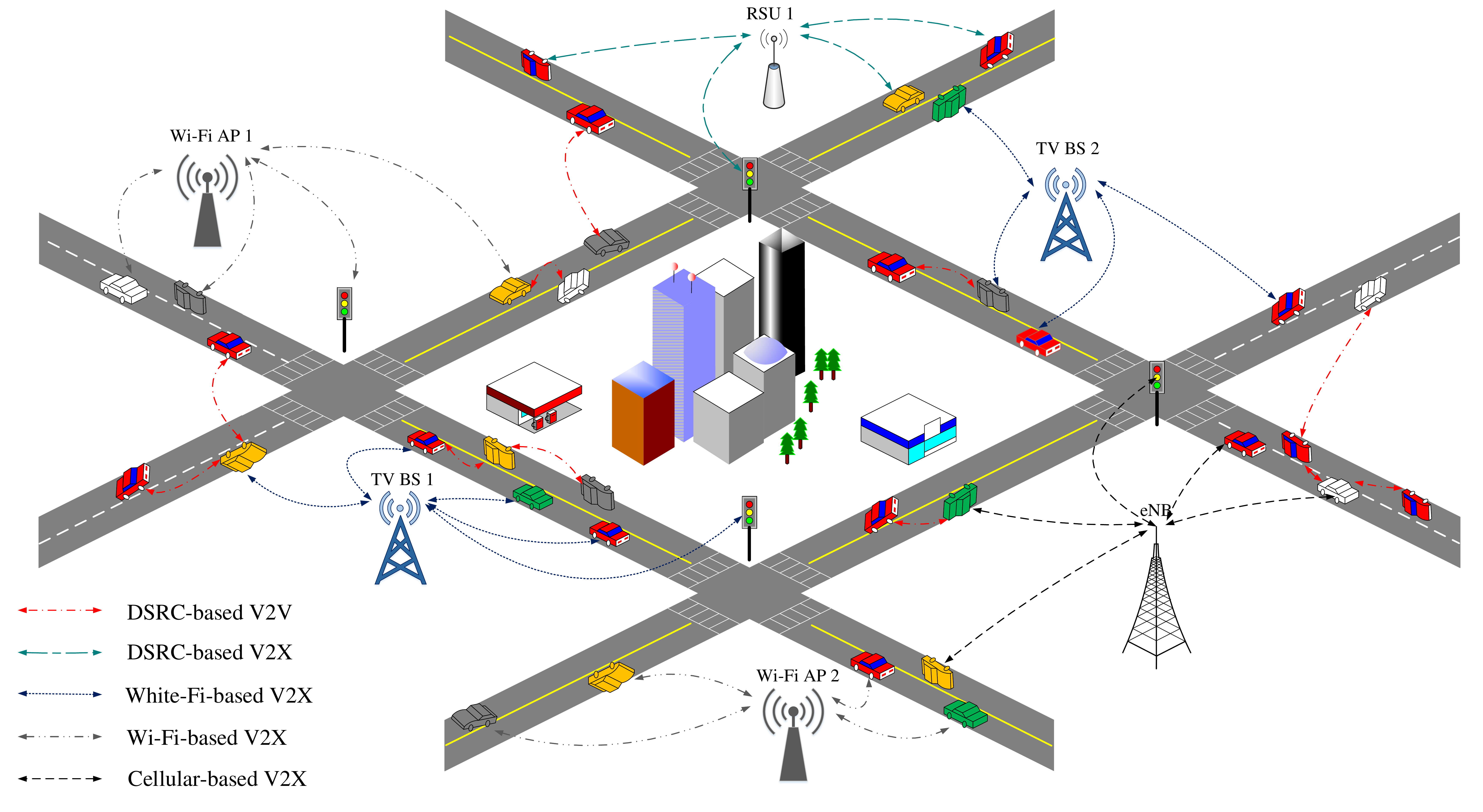}
\caption{Illustration of a heterogeneous MDVNET in an urban area}
\label{Fig.heterogeneousIDVNET}
\end{figure}

\textbf{\emph{Multiple technologies interworking:}} It has been shown that single technology applied in vehicular networks always has its own limitations as mentioned before. Thus, instead of establishing the \emph{homogeneous MDVNETs} that support inter-vehicle communications by a single wireless communication technology, the aforementioned advantages and shortcomings of DSRC, cellular, Wi-Fi, and White-Fi have motivated the works on establishing heterogeneous MDVNETs \cite{Lee2014Khop, zheng2015Heterogeneous, Abboud2016Interworking}. In \emph{heterogeneous MDVNETs}, inter-vehicles communications are supported by at least two kinds of wireless communication technologies, example of a heterogeneous MDVNET in an urban area is illustrated in Figure \ref{Fig.heterogeneousIDVNET}, where V2X refers to vehicle-to-everything communications, including V2V, V2I, and V2P communications. A typical heterogeneous MDVNET is the interworking of the DSRC and cellular technologies, where cellular-based communications can act as 1) a backup for traffic data when DSRC-based multihop links are disconnected in sparse vehicles situations, 2) a long-range Internet access method, and 3) a powerful backbone network for control message dissemination \cite{Abboud2016Interworking}. Other types of heterogeneous MDVNETs include the interworking of the Wi-Fi and cellular technologies \cite{taleb2015vecos}, interworking of the DSRC and White-Fi technology \cite{lim2014interplay}, interworking of the DSRC, Wi-Fi, and cellular technologies \cite{mojela2013use}. Even though multiple technologies interworking can make best use of the advantages and bypass the disadvantages of each single technology, how to select the applicable technology for each communication link and achieve seamless handoff among different technologies are still challenging.

\subsection{Routing Protocols}
\label{subsec:RP}

\begin{figure}[htbp]
\centering
\subfigure[Unicast]{
\label{fig:Unicast}
\includegraphics[width=0.480\textwidth]{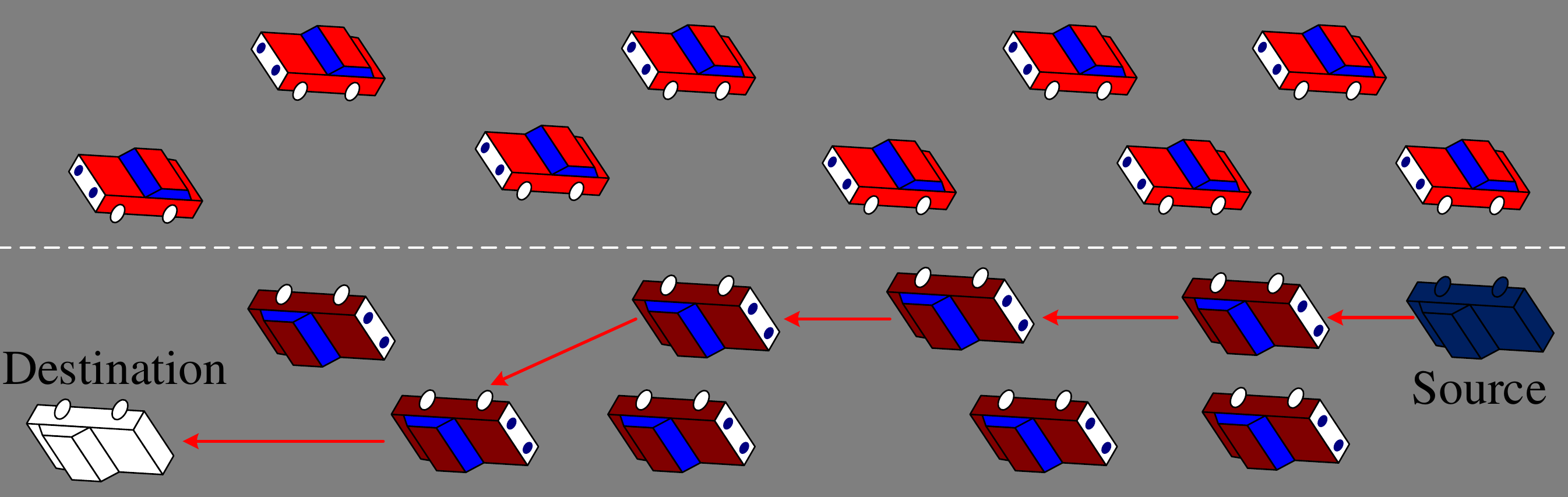}}
\hspace{0cm}
\subfigure[Multicast]{
\label{fig:Multicast}
\includegraphics[width=0.480\textwidth]{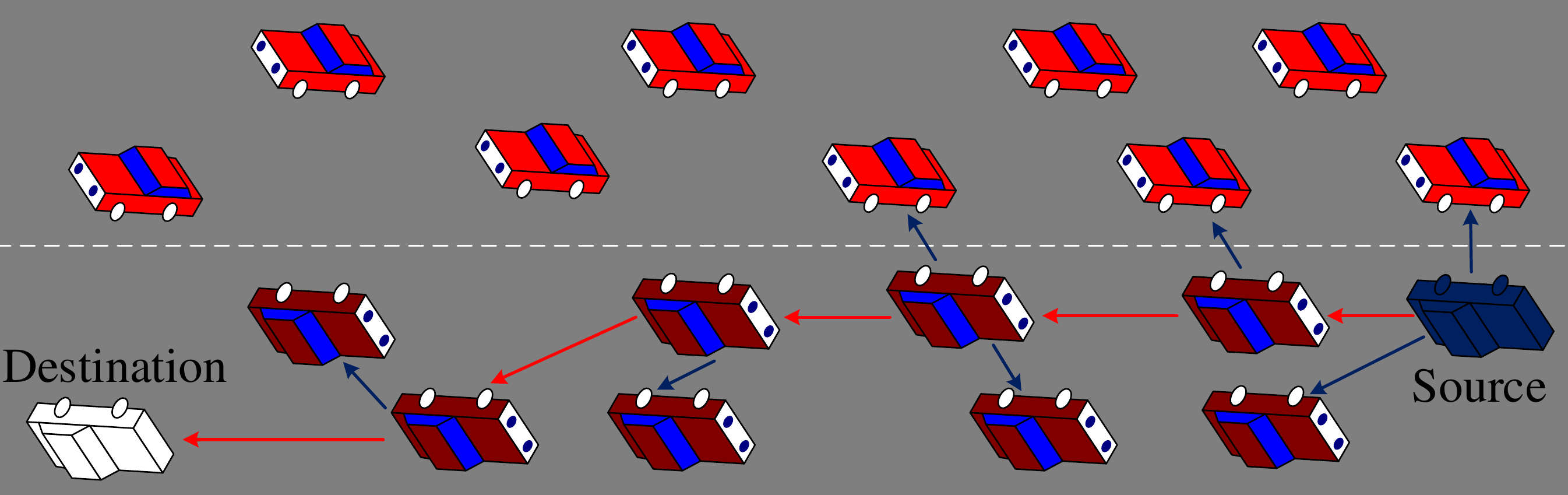}}
\hspace{0cm}
\subfigure[Broadcast]{
\label{fig:Broadcast}
\includegraphics[width=0.480\textwidth]{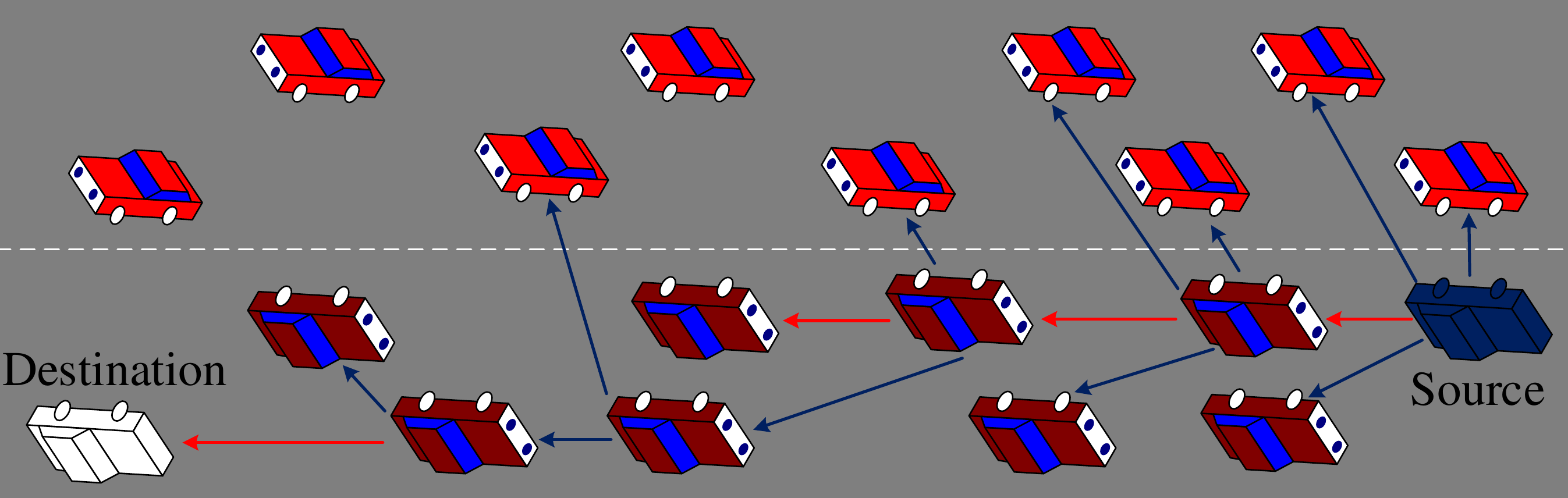}}
\hspace{0cm}
\subfigure[Cluster-based]{
\label{fig:Cluster}
\includegraphics[width=0.480\textwidth]{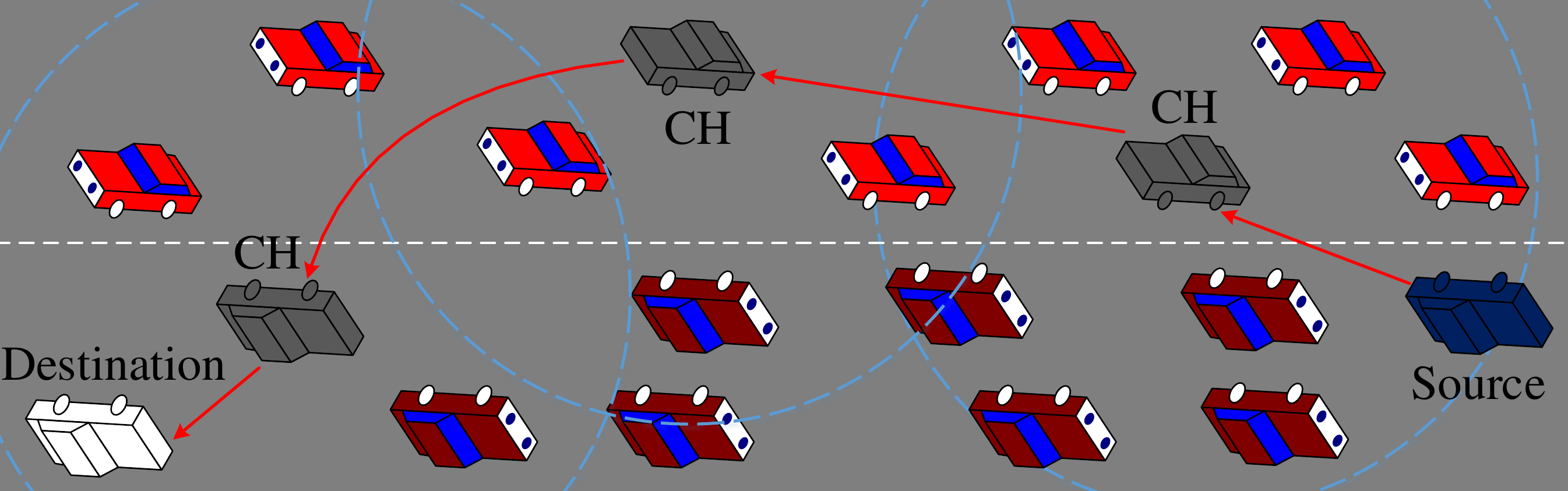}}
\caption{Illustration of four categories of routing protocol in MDVNETs}
\label{fig:Routingtypes}
\end{figure}

A routing protocol specifies the way that two communication entities exchange information with each other. It includes establishing a route, deciding to forward the information, and acting in maintaining the route or recovering from route failure. For wireless communications, the main purpose of routing protocols is to reduce communication time while using minimum amount of network resources (devices and spectrum resource). To support various applications of vehicular networks, routing protocols for MDVNETs have been studied. Based on the transmission strategies used for routing in MDVNETs, we classify the routing protocols into four categories: unicast, multicast, broadcast, and cluster-based routing protocols, as shown in Figure \ref{fig:Routingtypes}.

\textbf{\emph{Unicast Routing Protocol:}} It refers to a one-to-one transmission from one communication entity to another. In MDVNETs, the main goal of unicast is to transmit packets from a single source vehicle to another single destination vehicle via single/multi hops wireless communications, by either using a ``hop-by-hop" mechanism or ``store-and-forward" one \cite{cheng2015routing}. The main difference between these two mechanisms is whether the intermediate vehicles forward the received packets immediately or carry them until the corresponding routing algorithm makes a forward decision \cite{zeng2015improved}. In the existing research, unicast routing protocols have been investigated in two ways: greedy and opportunistic.

In greedy unicast routing protocols, the source vehicle forwards the packets to its outermost neighbors (the next hop intermediate vehicles), and then the intermediate vehicles forward these packets to its outermost neighbors (the second hop intermediate vehicles) until these packets are received by the destination vehicle \cite{zhu2015stochastic}. That is, the forward decisions in the greedy-based unicast routing protocols are made based on the geographic information of the vehicles. It has been shown in \cite{zhu2015stochastic} that the greedy-based unicast routing protocols are well working in some simple communication scenarios when ``hop-by-hop" mechanism is used, such as for vehicles in straight traffic roads. However, for urban areas that always have lots of intersections, greedy store-and-forward routing protocols are more suitable to support delay-tolerant non-safety applications \cite{jerbi2009towards}. For example, by exploiting the availability of map information, they can reduce the communication delay through choosing the next hop based on the information about physical location, velocity, direction \cite{zhao2008vadd}, dynamic traffic density, and curve-metric distance to the destination \cite{jerbi2009towards}, etc.

Opportunistic unicast routing protocols support wireless networks to operate in the scenario with frequent disconnections, such as vehicular networks \cite{spyropoulos2009routing}. In opportunistic unicast routing protocols, data packets from the source are delivered to the destination opportunistically, where 1) the intermediate vehicles should have ability to store-and-carry the received packet and perform in the ``store-and-forward" manner; 2) forward decisions are made independently for vehicles with different regions, to deliver the packets eventually with little or no location information of the destination; and 3) multiple copies of the source packets may be transmitted in parallel to increase the probability of at least one copy of the packets being delivered. At the beginning, most opportunistic unicast routing protocols are proposed for MANETs with homogeneous settings of mobile nodes \cite{spyropoulos2008efficient}. In order to design suitable opportunistic unicast routing protocols for networks with heterogeneous settings, such as vehicles with different speeds, relays (intermediate vehicles) are selected according to some utility functions in \cite{spyropoulos2009routing}. Moreover, a street-centric opportunistic routing protocol has been proposed in \cite{zhang2016street}, where the signal fading and mobility of vehicles have been considered when choosing the intermediate vehicles. The major challenges in designing unicast routing protocols include how to efficiently share these essential information, reduce the communication delay and packet losses, and deal with the conflict of routing when there are a large number of unicast routing requests.

\textbf{\emph{Multicast Routing Protocol:}} It is a one-to-many or many-to-many group communication where packets from the source are addressed to a group of destinations simultaneously. It has been demonstrated in \cite{bachir2003multicast, zhang2016geomobcon} that multicast routing protocols are essential to inter-vehicle communications among a group of vehicles in some situations, such as roadblocks, traffic congestion, calamities, and geographic advertising. The most classical and familiar multicast routing protocols are geocast, which are also special forms of multicast used for MANETs, where the group of destinations are identified by their geographical locations. In \cite{bachir2003multicast}, an Inter-Vehicle Geocast (IVG) protocol has been proposed to inform all vehicles about any danger in a highway, where the vehicle's location, speed, and driving direction are taken into account. The geocast protocol can be improved by considering more and different information, such as mobility and connection information of vehicles \cite{zhang2016geomobcon}, and vehicle trajectories \cite{jiang2015tmc}. The main goal is to address the challenges caused by network disconnection, mobility uncertainty, and redundant packets, and therefore design more efficient multicast protocols. Another important issue in designing multicast routing protocols is the security, namely, how to make sure only the authorized vehicles can be involved in the related communications and provide the confidentiality of the communications among an authorized group of vehicles. To address this issue, group key management schemes have been proposed with or without the assistance of infrastructure \cite{park2011rsu}.

\textbf{\emph{Broadcast Routing Protocol:}} It uses a one-to-all communication method to transfer a message from a single source to all receivers simultaneously. For vehicular networks, broadcast is an important routing method, which is usually used to discover nearby neighbors, propagate useful traffic information to other vehicles to support safety-related and cooperative driving applications, and disseminate the aforementioned essential information for unicast or multicast methods. The availability and practicability of broadcast are beyond doubt. Nevertheless, two major problems should be considered when designing a broadcast protocol for vehicular networks, especially for DSRC-based vehicular networks: broadcast storm and disconnected network problem.

A broadcast storm usually occurs in the communication environments with redundant broadcasting packets and high density vehicles, especially when multiple vehicles attempt to transmit packets simultaneously. When a broadcast storm is occurring, transmission among neighboring vehicles may experience frequent channel resource contention and packets collisions. There are a number of works to address this issue for vehicular networks. In \cite{wisitpongphan2007broadcast}, three distributed broadcast suppression techniques have been proposed, where a vehicle rebroadcasts the received broadcasting packets with a probability less than or equal to one. In these three techniques, the rebroadcast probability for each vehicle depends on the distance between this vehicle and the sender, that is, only global positioning system (GPS) information or packet received signal strength (RSS) information from vehicles within one-hop area is required. Besides the distance information, other information has been also used to select rebroadcast vehicles, such as local topology information \cite{tonguz2010dv} and spatial distribution information \cite{slavik2013spatial}. Furthermore, advanced techniques are applied in recent proposed broadcast protocols. Such as in \cite{wu2015efficient}, network coding is used to reduce the number of transmissions, and therefore reduce the redundant broadcasts.

The disconnected network problem is one of the major problems for each type of routing protocol. Similar to unicast and multicast routing protocols, store-and-forward method is usually applied in broadcast protocols to address the disconnected network problem. For example, by integrating rebroadcast vehicle selection and store-and-forward, a distributed multihop broadcast protocol has been proposed in \cite{tonguz2010dv}, which can not only handle both the broadcast storm and disconnected network problem but also operate in different traffic environments, including extreme scenarios with dense or sparse vehicles.

\textbf{\emph{Cluster-based Routing Protocol:}} It is a routing method by grouping vehicles into different sets (clusters) according to some rules, selecting a vehicle from each set to be a cluster head (CH), and the rest are called cluster members (CMs). Note that, unicast, multicast, and broadcast are three types of routing protocols that can be used to describe all routing methods in vehicular networks. More precisely, cluster-based routing protocol is a combination of unicast, multicast, and broadcast rather than regarding it as a new type of routing protocol. For example, the communication mode in a cluster can be a CM unicasts packets to its CH, a CH broadcasts packets to all its members and multicasts to the road infrastructures or other CHs, and the road infrastructures unicast packets to one CH or multicast packets to CHs within its coverage \cite{cooper2017comparative}.

Existing research results have indicated that clustering can improve the scalability and reliability of routing protocol in vehicular networks, by grouping vehicles based on information about relative velocity and spatial distribution. Due to the hierarchical characteristics of the cluster-based routing protocol, it is especially suitable for mulithop vehicular networks with large scale and heterogeneous communication technologies \cite{wang2008position, benslimane2011dynamic, ucar2016multihop}. In heterogeneous MDVNETs, cluster-based routing protocol can make the selection process of communication technologies simplified and effective. For example, in a cluster-based heterogeneous MDVNETs with the interworking of DSRC and cellular technologies, a simple way to select communication technologies for each links is that the CMs communicate with the CH by using DSRC and the CHs communicate with the eNB by using cellular technologies \cite{ucar2016multihop}. This way can minimize the number of vehicles communicating by using cellular technologies, and therefore reduce the cost of vehicular networks and the overload of cellular networks.

Despite the advantages of cluster-based routing protocols, an important issue should be considered when designing a cluster-based routing protocol, i.e., the stability of cluster membership. The cluster member stability, usually represented by residence times of cluster, relates to cluster head selection mechanism and can be analyzed by stochastic mobility model \cite{abboud2016stochastic}. For example, in order to improve the stability of cluster membership, lots of information, such as geographic position information of vehicles \cite{wang2008position}, the relative space position relations between vehicles and the center of the cluster \cite{benslimane2011dynamic}, and the relative velocity between a vehicle with its neighboring vehicles \cite{ucar2016multihop}, should be considered when selecting CHs. Table \ref{table:routing} summarizes unicast, multicast, broadcast, and cluster-based routing protocols in terms of specific examples, elementary information, mobility model, communication model, application type, and hierarchical topology.

\begin{table*}[htbp]
\begin{center}
\setlength{\belowcaptionskip}{5pt}
\caption{\label{table:routing} Comparisons of routing protocols in MDVNETs}
\centering
\begin{tabu} to 1\textwidth{X[c]|X[1.5,c]|X[2.5,c]|X[1.4,c]|X[1.6,c]|X[2,c]|X[1.5,c]}
    \toprule
Routing type & Specific examples & Elementary information & Mobility model & Communication model & Application type & Hierarchical topology \\
\cline{1-7}
    \multirow{4}{*}{Unicast}
    & GyTAR {\cite{jerbi2009towards}} & Traffic density \& Curvemetric distance  &  City  &  V2V  &   Safety \& Comfort applications  &  No \\
    \cline{2-7}
    & VADD {\cite{zhao2008vadd}} & Location \& Velocity \& Direction & Highway \& City & V2V \& V2I & Delay tolerant & No \\
    \cline{2-7}
    & {\cite{spyropoulos2009routing}} & NA & Highway \& City & V2V \& V2I \& V2S & Delay tolerant & No \\
    \cline{2-7}
    & ETCoP{\cite{zhang2016street}} & Mobility \& Local topology  & City & V2V \& V2I \& V2S & Delay tolerant & No \\
\cline{1-7}
 \multirow{3}{*}{Multicast}
    & IVG {\cite{bachir2003multicast}} & Location \& Velocity \& Direction  &  Highway  &  V2V  &  Safety &  No \\
    \cline{2-7}
    & GeoMobCon {\cite{zhang2016geomobcon}} & Mobility \& Contact  &  Highway  &  V2V  &  Delay-tolerant &  No \\
    \cline{2-7}
    & TMC {\cite{jiang2015tmc}} & Trajectories  &  City  &  V2V  &  For public vehicles &  No \\
\cline{1-7}
 \multirow{3}{*}{Broadcast}
    & DV-CAST {\cite{tonguz2010dv}} & Local location \& velocity &  Highway &  V2V  &  Safety &  No \\
    \cline{2-7}
     & DADCQ {\cite{slavik2013spatial}} & Location  &  Highway \& City &  V2V  & NA &  No \\
     \cline{2-7}
     & {\cite{wu2015efficient}} & Location \& Velocity &  Highway \& City &  V2V  & NA &  No \\
\cline{1-7}
 \multirow{3}{*}{Cluster}
    & VMaSC–LTE {\cite{ucar2016multihop}} & Relative velocity &  Highway  &  V2V \& V2I &  Safety &  Yes \\
    \cline{2-7}
    & PPC {\cite{wang2008position}} & Location \& Vehicle priority &  Highway  &  V2V &  NA &  Yes \\
    \cline{2-7}
    & \cite{benslimane2011dynamic} & Local location \& direction &  Highway \& City  &  V2V \& V2I &  NA &  Yes \\
    \bottomrule
\end{tabu}
\end{center}
\end{table*}

\subsection{Handoff Strategies}
\label{subsec:HS}

In MDVNETs, handoff is a major issue for efficient vehicular communications. Vehicles often move in and out of the communication ranges of other vehicles and infrastructures, resulting in frequently disconnected V2V and V2I communication links and dynamically changing network topology. Handoff strategies aim to provide a seamless communication for vehicles in MDVNETs while reducing financial cost, handoff latency, and packet loss, etc, and have attracted lots of attentions in the past few years. There are two types of handoff strategies: horizontal handoff and vertical handoff, depending on whether the same or different wireless access technologies are used.

\textbf{\emph{Horizontal handoff:}} It is used for transferring a data transmission session from one point of attachment to another, where the same wireless access technology is applied in both points of attachments \cite{Abboud2016Interworking}. There are three scenarios where horizontal handoff is required in MDVNETs, i.e.,
\begin{itemize}
\item[1)] for V2V communications in vehicular ad hoc networks, horizontal handoff is required for data transfer when the neighboring vehicles change, and is performed by rerouting to construct a new routing path to the destination vehicle \cite{zhu2011mobility};
\item[2)] for D2D-based V2V communications in cellular-based MDVNETs, horizontal handoff is required to provide continuous communication services when the ProSe-enabled vehicle moves across the cell boundary \cite{chen2015handover}; and
\item[3)] for V2I communications in infrastructure-based homogeneous MDVNETs, horizontal handoff is required to transfer a vehicle's data transmission session, such as Internet video streaming session, from one infrastructure to another when the vehicle moves between both infrastructures' coverage ranges \cite{Abboud2016Interworking, chung2011time}. For example, the handover controller proposed in \cite{chung2011time} combines the centralized and distributed control mechanism to improve the efficiency of channel resources and support the handover operation for V2I communications.
\end{itemize}

\textbf{\emph{Vertical handoff:}} It is used for transferring a data transmission session from one point of attachment to another, where two different wireless access technologies are applied in these two points of attachments \cite{Abboud2016Interworking}. Different from horizontal handoff, vertical handoff only occurs in heterogeneous MDVNETs and is required to maintain the communication connection for data transfer when a vehicle moves out of the coverage area of one type of network to another, or is an optional and efficient method to benefit the users or networks when a vehicle moves in an overlapping area of two or more types of networks. Since the overlapping areas of different networks are very common in heterogeneous MDVNETs, vertical handoff as an optional method is mainly discussed in this context. Vertical handoff could be triggered by different reasons, such as user-centric handoff triggers, network-centric ones, and their combination.

The user-centric handoff triggers include:
\begin{itemize}
\item[1)] Quality of Service (QoS), the QoS for each vehicle user is generally based on common communication performance metrics, such as the end-to-end delay, throughput, outage probability, and packet loss. Different networks can provide vehicular users with different QoS. For example, when vehicles making handoff decision with probability based on the maximum transmission rate, their throughput can be improved. In \cite{ryu2014enhanced}, an enhanced mobile IPv6 fast handovers based on network mobility (NEMO) has been proposed, which can significantly mitigate the tunneling burden by delivering parts of packets via paths between the home agent and the new access routers without any tunneling and reduce the handover latency through performing the registration to the home agent in advance. Furthermore, an efficient proxy mobile IPv6 (E-PMIPv6) handoff scheme has been proposed in \cite{bi2016efficient} to guarantee session continuity in urban scenario, which eliminates packet loss and reduces handoff latency and signaling overhand.
\item[2)] Financial cost, applying different technologies results in different financial cost for vehicle users, such as subscription fees required by cellular technology and fees for RSUs placement and management. Vehicles can calculate handoff probability based on the costs of different networks. Therefore, network cost normalization method is important to this trigger. For example, in \cite{shafiee2011optimal}, distributed optimal vehicle handoff (VHO) algorithms have been proposed to minimize the packet end-to-end delay and the cost of data traffic transmission.
\end{itemize}

The network-centric handoff triggers include: 1) network throughput maximization, 2) fairness balancing, and 3) load balancing. In \cite{xu2015fast}, a fast network selection scheme, where vehicles select a access network according to a coalition formation game method, has been proposed to balance the throughput and fairness. In \cite{lee2009vertical}, by optimizing a combined cost function of the battery lifetime of the mobile nodes and load balancing over the infrastructures, a VHO decision algorithm has been proposed to balance the overall load among all infrastructures and maximize the collective battery lifetime.

In network-centric handoff triggers, handoff decisions are usually made based on the centralized calculating results and, therefore, abundant computing ability and data storage resources are important. Even though infrastructures in heterogeneous MDVNETs are strong enough to handle data computing and storing, cloud computing has been utilized to provide any time and anywhere data processing and storing for some network-centric handoff triggers, such as the fast cloud-based network selection scheme in \cite{xu2015fast}. Moreover, software defined network controller is also utilized in heterogeneous MDVNETs, where the handover strategy is combined with OpenFlow to make smoother handover decisions \cite{he2016cost}. Security is another important issue when doing handover since the security key needs to be changed for different connections. Thus, some works have been done to address this issue. For example, a QoS-aware distributed security architecture has been proposed in \cite{rengaraju2014qos}, where the elliptic curve Diffie-Hellman protocol is used to ensure the security strength of the proposed architecture. In a multihop-authenticated proxy mobile IP scheme proposed in \cite{cespedes2013multihop}, the concept of symmetric polynomials has been used for key generation at the mobile and relay routers for authentication.

\section{Automated Driving Vehicular Networks}
\label{sec:Platoon}

\begin{figure*}[htbp]
\centering
\includegraphics[height=1.6 in]{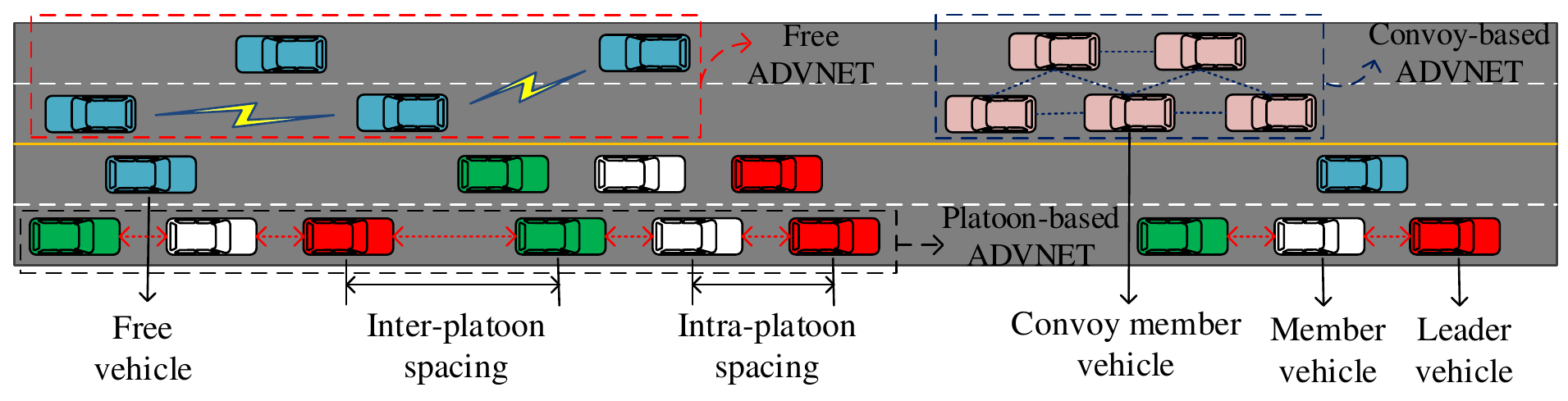}
\caption{Illustration of normal and platoon-based ADVNETs}
\label{Fig.ADVNETs}
\end{figure*}

Autonomous vehicle opens the door for fleet management and coordinative driving. According to the traffic management strategy applied in ADVNETs, we classify ADVNETs into three categories: free ADVNETs, convoy-based ADVNETs, and platoon-based ADVNETs, as shown as shown in Figure \ref{Fig.ADVNETs}. In free ADVNETs, AVs move independently under the control of the computerized autopilots \cite{moradi2016hybrid}. In convoy-based ADVNETs, AVs that move on multiple lanes and with the same direction are grouped into a convoy, where lateral and longitudinal controls are conducted by all AVs in the convoy on a distributed basis to maintain the pre-set inter-vehicle spacing and align velocity. Convoy is an extension of the idea of platoon \cite{de2014network}. In the platoon-based ADVNETs, with the centralized control of the leader vehicle (the first vehicle of the platoon), AVs move cooperatively with the same steady velocity and keep a steady inter-vehicle spacing one after another. Platoon-based management strategy can not only reduce the energy consumption and exhaust emission by minimizing air drag due to the streamlining of AVs, but also increase the driving safety by cooperative driving among vehicles \cite{jia2016survey}. Note that, platoon-based structure can be also considered in MDVNETs \cite{biswas2006vehicle}, where manually driving vehicles may spontaneously form a platoon, that is, one vehicle follows another one for easy driving. However, since the probability that several manually driving vehicles move spontaneously with a platoon structure is too small, the existing works related to platoon-based MDVNETs are considered as cluster-based MDVNETs, where the first vehicle of one platoon, i.e., the leader vehicle, acts as the CH, and others, i.e., member vehicles, are CMs.

Different from MDVNETs, the impacts of the high mobility and complicated communication environments on the automated driving vehicular networks are more significant due to the following reasons:
\begin{itemize}
\item[1)] nodes in ADVNETs, i.e., AVs, are self-driving without human's actions and interventions and have more stringent delay constraints and higher reliable packet delivery requirement to ensure the safety of AVs \cite{hobert2015enhancements}, especially for intersection environments;
\item[2)] more types of safety-related information are required for AVs' applications, resulting in challenging data load in ADVNETs; and
\item[3)] management and allocation of computing resources are more complicated, where both sensory and communication information should be processed for AVs.
\end{itemize}
In order to improve the safe navigation of AVs and efficient of information exchange, research and testing on ADVNETs have attracted a great deal of attention. In this section, existing works related to ADVNETs will be discussed. Specifically, we first present the different communication structures designed for ADVNETs, and then, we introduce the related wireless communication technologies used for sharing these information.

\subsection{Communication Structures}
\label{subsec:CIS}

Communication structures indicate the communication pair or group of AVs for sharing different types of information. For different types of ADVNETs, different communication structures are designed.

In free ADVNETs, one of the main communication structures is broadcast-based V2V communication, i.e., each AV broadcasts its information to the neighboring vehicles \cite{naumann1998managing, han2015styled, moradi2016hybrid}. For example, in \cite{naumann1998managing}, a collision-avoidance algorithm is proposed to prevent collision in intersections, where an AV broadcasts information about collision to AVs within its communication range once it moves in a critical region, and therefore avoiding endless waiting in the intersection. To avoid collision in usual road environment, a broadcast-based communication structure is designed with adjusting the communication range of each AV accordingly for different scenarios in \cite{moradi2016hybrid}. Another communication structure is V2I communication in infrastructure-based scenarios, where infrastructures, such as RSUs, are used for collecting and broadcasting information from or to AVs.

Convoy has been introduced by AutoNet \footnote{AutoNet system consists of infrastructure and cooperative vehicles, where the cooperative vehicles refer to the vehicles that are able to sense surrounding information and communicate with other cooperative vehicles for exchanging perception and commands to enable automated maneuvering.} 2030 in \cite{de2014network} to support cooperative driving among AVs on multiple lanes. At least two kinds of communication structures are considered in convoy-based ADVNETs, i.e., broadcast-based V2V communication structure for sharing information to enhance the cooperative maneuvering and V2I communication structure to improve efficiency of traffic management \cite{hobert2015enhancements}.

For platoon-based ADVNETs, in addition to ensuring safe driving of each AV, another major challenge is maintaining the string stability of platoons, i.e., ensuring that the inter-platoon spacing error and intra-platoon spacing error do not amplify upstream from AV to AV and from platoon to platoon. In addition to the information mentioned in Subsection \ref{subsec:VNA}, it has been indicated that velocity and acceleration information of the leader and preceding vehicles are required to share with all member vehicles and the following vehicle, respectively, to maintain string stability of the platoon \cite{Rajamani2011}. Thus, more communication structures are considered in platoon-based ADVNETs, including:
\begin{itemize}
\item[1)] intra-platoon communication structure \cite{michaud2006coordinated}, used for sharing information among AVs within the same platoon, can be further classified into three ones: leader-to-member (or member-to-leader) communication structure for sharing information between the leader vehicle and one of member vehicles \cite{xu2014communication}, two-adjacent communication structure for sharing velocity and acceleration information \cite{Rajamani2011, peng2017performance}, and platoon-based multicast communication structure for improving the communication efficiency.
\item[2)] inter-platoon communication structure is designed for sharing information among AVs within different platoons, such as sharing collision information from one platoon to its following platoons.
\end{itemize}
Similar to the free ADVNETs, infrastructures would play the control role either for network or platoons in most of platoon-based ADVNETs. Hence, in the infrastructure-based environments, infrastructures are usually considered to relay information for AVs \cite{raza1997vehicle, peng2017resource}.

\subsection{Technologies for ADVNETs}
\label{subsec:ACS}

Several wireless communication technologies have been considered to be applied in ADVNETs, including the ones that have been widely applied in MDVNETs, i.e., DSRC \cite{kato2002vehicle, moradi2016hybrid}, cellular \cite{Campolo2017-Better}, Wi-Fi \cite{milanes2010controller}, and White-Fi \cite{zhang2011cognitive} and other short-range wireless communication technologies, such as infrared and VLC. Among them, most of the long-range communications in ADVNETs and communications in the free ADVNETs are usually supported by the DSRC and cellular technologies, such as in \cite{kato2002vehicle, moradi2016hybrid, Campolo2017-Better}. Since the advantages, disadvantages, and challenges of DSRC, cellular, Wi-Fi, and White-Fi have been introduced in Section \ref{sec:Indivi}, we will focus on the infrared and VLC in this subsection.

\textbf{\emph{Infrared (IR):}} It is one of the earliest technologies that uses light invisible to human, i.e., infrared light, to enable wireless communications on a line-of-sight (LOS) basis. It has been indicated in \cite{fernandes2012platooning, aoki1996inter} that infrared communication has high directionality, high confidentiality, and harmlessness. Considering the roads environments are usually in line, communications among AVs can be achieved by an inexpensive infrared laser. Furthermore, the wavelength of infrared communications is much shorter than other wireless communications while it is with a potential of 1 Gbps data communication speed. Therefore, it is optimal for sharing large-volume data, such as multimedia entertainment information among AVs. Infrared lights have been used to measure the intra-platoon spacing to assist the gap control in \cite{aoki1996inter}. However, since the infrared light can only carry shorter than around 10 meters and do not penetrate obstacles, infrared communications can only be used in some special scenarios, such as for two-adjacent communication structure in platoon-based ADVNETs \cite{aoki1996inter}. Furthermore, weather conditions are another challenge for infrared communications. Thus, in order to take the advantages while addressing related challenges, infrared communications are always combined with other wireless communication technologies, such as DSRC, to support AVs' applications \cite{fernandes2012platooning}.

\textbf{\emph{Visible light communications (VLC):}} It is a type of communication through visible light. Compared with other types of wireless communications, VLC has the following special importance advantages:
\begin{itemize}
\item[1)] 360 Terahertz of license free bandwidth within visible light spectrum can be tapped for wireless communications while avoiding the interference from radio frequency (RF) signals. Therefore, it can address the scarcity of current RF spectrum to support the ever increasing mobile data traffic demand. For example, in \cite{segata2016platooning}, VLC is used to backup DSRC communications to reduce the packet loss resulted from congestion in scenarios with high density vehicles;
\item[2)] Existing research about VLC has shown that a very high data communication speed can be achieved by VLC, i.e., multiple Gbps in research and around 100 Mbps in IEEE 802.15.7 standard; and
\item[3)] By reusing existing lighting infrastructure to support communication, VLC can be deployed easily and inexpensively \cite{pathak2015visible}.
\end{itemize}
The above advantages have motivated work on tapping VLC for ADVNETs. For example, the feasibility of using VLC for sharing information among platoon members to support platoon control has been studied in \cite{abualhoul2013platooning}. Similar to infrared communications, VLC signals cannot penetrate through most obstacles. Therefore, it can be only used for communications between two vehicles with LOS.

It should be noted that IR and VLC technologies can not only be used to support communications in ADVNETs but also in MDVNETs. For example, IR has been used in tolling applications \cite{shieh2017vehicle}. Based on the light emitting diode (LED) headlamp, a demonstration system that uses VLC to support V2V communications has been proposed in \cite{yoo2016demonstration}.

\section{Challenges and opportunities}
\label{sec:CAO}

Even though there has been a large amount of existing research for vehicular networks and many industrial products can be used to support vehicular communications, achieving efficient vehicular communications to support practical ITS applications still faces many challenges. In this section, we will discuss the major challenges and opportunities.

\subsection{Heterogeneous Driving Vehicular Networks}
\label{subsec:HDVN}

With the rapid development of autonomous vehicles, AVs will gradually prevail on the roads in the near future, resulting in scenarios where AVs and manually driving vehicles move on the roads simultaneously. In order to achieve information sharing among manually driving and autonomous vehicles, a heterogeneous driving vehicular network (HDVNET) is needed and its realization faces the following challenges.

It has been indicated that cooperative driving with ADVNETs is a promising technology to improve safety driving of AVs, especially in intersection scenarios. For example, through sharing real-time positions, velocity, and desired driving lane information among AVs within the same platoon, cooperative based safety driving patterns can be designed to help AVs to cross the intersection without traffic lights safely in \cite{li2006cooperative}. However, how to help AVs and manually driving vehicles to cross intersections safely remains to be unsolved. In such scenarios, manually driving vehicles have to be controlled by drivers, resulting in human factors, and therefore, challenging the cooperative driving.  Thus, designing an efficient and suitable communication scheme, i.e., scheduling which kind of information should be shared by which vehicle, to achieve efficient cooperative driving is desired for some scenarios such as intersections.

In HDVNETs, AVs and manually driving vehicles have different information requirements both in message types and QoS requirements, which result in complex message structure. In \cite{Liu2017-infrastructure}, a novel beacon scheduling algorithm has been proposed to guarantee the reliable and timely dissemination of two types of beacon messages in HDVNETs, i.e., event-driven safety messages for manually driving vehicles and periodic beacon messages for cooperative driving of AVs. However, in most HDVNETs scenarios, message types are more complex than the one that has been considered in \cite{Liu2017-infrastructure}. For example, when a special infotainment service (e.g, video) is required for vehicles in platoon-based HDVNETs, event-driven safety messages, periodic messages, and video information need to be shared to guarantee the string stability of platoon while ensuring the QoS requirement of this infotainment service. Hence, different efficient message dissemination schemes are required for different HDVNETs scenarios.

\subsection{Security Issue}
\label{subsec:SI}

In addition to the security issue discussed in Subsection \ref{subsec:HS} for MDVNETs \cite{lu2012dynamic}, there are more security related concerns need to be considered when cooperative driving is considered in ADVNETs and HDVNETs. Since the final driving decisions in cooperative driving are made based on the shared information, security attacks either on the communication channel or sensor tampering significantly impact the safety driving \cite{amoozadeh2015security}. Thus, how to protect communications from message falsification, message eavesdropping, radio jamming, and tampering attacks is important to cooperative driving. However, the algorithms used for current wireless communication security may not work well for ADVNETs and HDVNETs since lower latency is required by communications in cooperative driving. In order to reduce the end-to-end delay while guaranteeing the information security, group authenticated algorithms can be considered, especially for communications among vehicles within the same platoon or convoy, as well as for vertical handoff in heterogeneous ADVNETs and HDVNETs. Moreover, for cluster-based MDVNETs and platoon/convoy-based ADVNETs, performing efficient mutual authentication in the formation stage of cluster/platoon/convoy and when a new vehicle is applying to join them, can reduce the impacts of authentication on information sharing in each cluster/platoon/convoy.

\subsection{Other Issues}
\label{subsec:OIs}

There are some other issues that should be considered in MDVNETs, ADVNETs, and HDVNETs. Firstly, mmWave and VLC communication technologies have been considered to support vehicular networks \cite{choi2016millimeter, segata2016platooning, yoo2016demonstration}. However, how to address the potential issues caused by high mobility and harsh weather conditions still needs further investigation and test. Moreover, in heterogeneous vehicular networks with one of or both of mmWave and VLC communication technologies, designing criterion for network selection between mmWave/VLC and other wireless communication technologies is not an easy work. Secondly, to achieve efficient cooperative driving, communication information management and cooperative control should be considered globally in ADVNETs and HDVNETs, which results in complicated allocation issue among storage and computing resources of each vehicle.

\section{Conclusions}
\label{sec:Conclu}

In this paper, we have provided a comprehensive survey in vehicular communications from the network layer perspective, including communications in MDVNETs and ADVNETs. In addition to providing infotainment services to passengers, MDVNETs can be used to improve the traffic safety for manually driving vehicles. ADVNETs are expected to support cooperative driving to further improve driving safety. Advantages and disadvantages of different communication technologies that can be applied in these two types of vehicular networks have been analyzed. Due to the high mobility and complicated communication environments, challenges confronted with vehicular communications are different when used to support different applications. Recent research works addressing these challenges have been discussed, including routing protocols for exchanging different messages, efficient handoff strategies to benefit the users or networks, and suitable communication structures according to the applied traffic management strategies. We have further identified several main under-explored issues in MDVNETs, ADVNETs, and HDVNETs, and provided potential solutions that can be considered.

\bibliographystyle{IEEEtran}
\bibliography{surveydata}

\ifCLASSOPTIONcaptionsoff
  \newpage
\fi

\end{document}